\documentclass[twocolumn,showpacs,amsmath,amssymb]{revtex4}

\usepackage{graphicx}
\usepackage{bm}
\begin{document}

\title{Interatomic van der Waals potential in the presence of
 a magneto-electric sphere}

\author{Hassan Safari}
\author{Dirk--Gunnar Welsch}

\affiliation{Theoretisch--Physikalisches Institut,
Friedrich--Schiller--Universit\"at Jena,
Max--Wien-Platz 1, D-07743 Jena, Germany}

\author{Ho Trung Dung}
\affiliation{Institute of Physics, Academy of
Sciences and Technology, 1 Mac Dinh Chi Street,
District 1, Ho Chi Minh city, Vietnam}

\author{Stefan Yoshi Buhmann}
\affiliation{Quantum Optics and Laser Science, Blackett Laboratory,
Imperial College London, Prince Consort Road,
London SW7 2BW, United Kingdom}

\date{\today}

\begin{abstract}
On the basis of a general formula obtained earlier via fourth-order
perturbation theory within the framework of macroscopic quantum
electrodynamics, the van der Waals potential between two neutral,
unpolarized, ground-state atoms in the presence of a homogeneous,
dispersing and absorbing magneto-electric sphere is studied. When the
radius of the sphere becomes sufficiently large, the result
asymptotically agrees with that for two atoms near a planar
interface. In the opposite limit of a very small sphere, the sphere
can effectively be regarded as being a third ground-state atom, and
the nonadditive three-atom van der Waals potential is  recovered. To
illustrate the effect of a sphere of arbitrary radius, numerical
results are presented for the triangular arrangement where the atoms
are at equidistance from the sphere, and for the linear arrangement
where the atoms and the sphere are aligned along a straight line.
As demonstrated,
the enhancement or reduction of the interaction potential in the
presence of purely 
electric or magnetic spheres
can be physically understood in terms of image charges.
\end{abstract}


\maketitle

\section{Introduction}
\label{Sec:Intro}
Electromagnetic and material ground-state fluctuations are well known
to produce observable effects such as dispersion forces between
atoms, between atoms and bodies, or between bodies \cite{rev2}. The
van der Waals (vdW) potential between two ground-state atoms in free
space was first calculated by London for small interatomic separation
(nonretarded limit), using perturbation theory to the leading (second)
order \cite{Lon}. In this limit, the result is an attractive potential
proportional to $r^{-6}$, where~$r$ denotes the interatomic distance.
The (nonretarded) London potential was extended to arbitrary distances
between the two atoms by Casimir and Polder \cite{CP}, using
fourth-order perturbation theory within the framework of full quantum
electrodynamics (QED). In particular for large interatomic separation
the potential was predicted to vary as $r^{-7}$.
Recently, the closely related Casimir interaction between two
magnetoelectric spheres has been studied by means of a scattering
method~\cite{Emig}, where the inclusion of higher-order multipoles
have been shown to lead to corrections of the Casimir--Polder result.
In the three-atom case, a non-additive term prevents the potential
from just being the sum of three pairwise contributions.
This three-atom vdW potential was calculated first in the
nonretarded limit by pursuing the perturbation calculation to the
third order \cite{Ax}, and then for arbitrary interatomic distances,
by using sixth-order perturbation calculation \cite{Aub}. Later on,
a general formula for the non-additive $N$-atom vdW potential was
derived by summing up the response of each atom to the quantized
field caused by the other atoms \cite{Pow1} and by calculating the
difference in the zero-point energy of the electromagnetic field of a
large cavity with and without the atoms \cite{Pow2}.

The presence of macroscopic bodies modifies the fluctuation of the
electromagnetic field, and consequently, the interatomic vdW
interaction. A general formula expressing the vdW potential between
two ground-state atoms in the presence of electrically polarizable
bodies in terms of the Green tensor of the body-assisted
electromagnetic field was first obtained by means of linear response
theory \cite{Mc}, an later by treating the effect of the bodies
semiclassically \cite{Mah1}. An exact derivation of the formula based
on fourth-order perturbation theory within the framework of
macroscopic QED has been given recently \cite{h}, where both
electrically and magnetically polarizable bodies---referred to as
magneto-electric bodies---are explicitly taken into account, and a
generalization of the formula to the $N$-atom case has also been given
\cite{BSW}.
The two-atom potential in the presence of arbitrary magnetoelectric
bodies has recently been generalized to
atoms having both electric and magnetic polarizabilities 
using linear response theory
and was explicitly calculated for atoms embedded in a bulk
medium~\cite{Spag1}.
Various special cases such as two atoms placed between two perfectly
conducting plates \cite{Mah1}, in a bulk magneto-electric medium
\cite{Mah2,Spag2,Sin,h}, near a perfectly reflecting plate \cite{h},
and near a planar magneto-electric multilayer system \cite{h} have
been considered. Local-field corrections to the vdW potential that
appear when the atoms are embedded in optically dense media have also
been addressed \cite{agnes,tomas}.

On the experimental side, the vdW interaction between a single atom
and a body has been explored by means of detecting the intensity of an
atomic beam transmitted through a parallel-plate cavity \cite{suk};
direct force measurement using atomic-mirror techniques
\cite{land,moh}; measuring the intensity of a diffracted atomic beam
from a transmission-grating \cite{gri};
making use of quantum reflection from a solid 
surface 
at
nonretarded~\cite{Druzhinina} 
and retarded~\cite{Pasquini1,Pasquini2} atom--surface separations;
determining their effect on the collective oscillation frequency
of the magnetically trapped atoms~\cite{Obrecht}.
Observations of interatomic vdW interactions based on a determination
of the scattering cross sections in the atomic collisions between two
ground-state atomic beams \cite{chen}, between an atomic beam and the
atoms of a stationary target gas \cite{bru1,bru2}, and between a beam
of ground-state atoms and a beam of excited atoms \cite{mar} have been
reported.

As already mentioned, the theoretical studies of medium-assisted
interatomic vdW interactions have so far concentrated on bulk media
and infinitely extended planar (multi-layer) bodies Here we shall
consider the vdW interaction between two ground-state atoms located
near a finite-size body, namely, a sphere. With recent progress in
fabrication of metamaterials in mind, we allow for the sphere to
exhibit both electric and magnetic properties. Note that the
single-atom vdW potential in the presence of a sphere has been
investigated earlier \cite{st1}.

The paper is organized as follows. The basic formulas for calculating
the vdW potential between two atoms in the presence of an arbitrary
arrangement of magneto-electric bodies are summarized in
Sec.~\ref{general}. In Sec.~\ref{sphere}, the theory is applied
to the case of a magneto-electric sphere, and the limiting cases of
large and small sphere are considered. Detailed numerical results are
presented in Sec.~\ref{num}. Finally, the paper ends with a summary
in Sec.~\ref{summ}.


\section{Basic formulas}
\label{general}

Consider two neutral, unpolarized, ground state atoms $A$ and $B$
with spherically symmetric polarizabilities in the presence of an
arbitrary arrangement of dispersing and absorbing magneto-electric
bodies. The total force acting on the atoms can be derived from the
potential
\begin{equation}
\label{eq1}
U(\mathbf{r}_A , \mathbf{r}_B)=U_A( \mathbf{r}_A) +
U_B( \mathbf{r}_B)+U_{AB}( \mathbf{r}_A , \mathbf{r}_B)
\end{equation}
according to
\begin{equation}
\label{F}
\mathbf{F}_{A'} (\mathbf{r}_{A'})=
-\bm{\nabla}_{\mathbf{r}_{A'}}U( \mathbf{r}_A , \mathbf{r}_B),
\end{equation}
($A'$ $\!=$ $\!A,$ $\!B$), where $U_{A'}$ is the single-atom potential
\cite{st1}
\begin{equation}
\label{single}
U_{A'}(\mathbf{r}_{A'})=
 \frac{\hbar\mu_0}{2\pi}\int_0^\infty
 \mathrm{d}u\,u^2\alpha_{A'}(iu)\mathrm{Tr}{\bm  G}^{(1)}
 (\mathbf{r}_{A'},\mathbf{r}_{A'},iu),
 \end{equation}
and $U_{AB}$ is the two-atom interaction potential \cite{h}
\begin{multline}
\label{vdw}
U_{AB}(\mathbf{r}_A,\mathbf{r}_B) =
 -\frac{\hbar\mu_0^2}{2\pi}\int_0^\infty
 \mathrm{d}u\,u^4\alpha_A(iu)\alpha_B(iu)
\\
\times\mathrm{Tr}[{\bm  G}
 (\mathbf{r}_A,\mathbf{r}_B,iu)\!\cdot\!{\bm G}
 (\mathbf{r}_B,\mathbf{r}_A,iu)].
\end{multline}
In Eqs.~(\ref{single}) and (\ref{vdw}), ${\alpha}_{A'}$ is the
(lowest-order) polarizability of the atom $A'$
\begin{equation}
\label{p}
\alpha_{A'}(\omega)=\lim_{\eta\rightarrow 0^+}\frac{2}{3\hbar}\sum_k
 \frac{\omega^{k0}_{A'}
|\mathbf{d}^{k0}_{A'}|^2}{(\omega^{k0}_{A'})^2
 -\omega^2-i\eta\omega}\,,
\end{equation}
with $\omega^{k0}_{A'}$ and $\mathbf{d}^{k0}_{A'}$ being, respectively,
the transition frequency and transition electric dipole moment between
the $k$th excited state and the ground state of atom $A'$, and
$\bm{G}(\mathbf{r}, \mathbf{r'},\omega)$ is the classical Green tensor
obeying the differential equation
\begin{multline}
\label{eq6}
\bm{\nabla}\times\frac{1}{\mu(\mathbf{r},\omega)}\bm{\nabla}
  \times
\bm{G}
  (\mathbf{r}, \mathbf{r'}, \omega)
\\
-\frac{\omega^2}{c^2}\,\varepsilon(\mathbf{r},\omega)
  \bm{G}
  (\mathbf{r}, \mathbf{r'}, \omega)
  =\bm{\delta}(\mathbf{r}-\mathbf{r'})
\end{multline}
together with the boundary condition at infinity. Note that the
electromagnetic and geometric properties of the bodies are fully
incorporated in the Green tensor via the space- and frequency-dependent
permittivity $\varepsilon(\mathbf{r},\omega)$ and permeability
$\mu(\mathbf{r},\omega)$. When $\mathbf{r}$ and $\mathbf{r}'$
denote two positions in free space which can be connected without
crossing a body, then the Green tensor can be decomposed as
\begin{equation}
\label{Gparts}
\bm{G}(\mathbf{r},\mathbf{r}',\omega)=
 \bm{G}^{(0)}(\mathbf{r},\mathbf{r}',\omega)+
 \bm{G}^{(1)}(\mathbf{r},\mathbf{r}',\omega),
\end{equation}
where $\bm{G}^{(0)}$ is the free-space Green tensor which is obtained
from Eq.~(\ref{eq6}) by letting $\varepsilon(\mathbf{r},\omega)$ $\!=$
$\!\mu(\mathbf{r},\omega)$ $\!=$ $\!1$, and $\bm{G}^{(1)}$ is the
scattering part of the Green tensor.

In what follows we concentrate on the atom--atom interaction potential
$U_{AB}$ in the presence of a sphere (for the single-atom potentials
$U_A(\mathbf{r}_A)$ and $U_B(\mathbf{r}_B)$, see Ref.~\cite{st1}).
According to Eq.~(\ref{Gparts}), the potential
$U_{AB}(\mathbf{r}_A,\mathbf{r}_B)$ can be cast in the form
\begin{equation}
\label{uparts}
U_{AB}(\mathbf{r}_A,\mathbf{r}_B) =
 U^{(0)}(\mathbf{r}_A,\mathbf{r}_B)
 +U^{(b)}(\mathbf{r}_A,\mathbf{r}_B),
\end{equation}
where
\begin{multline}
\label{vdw0}
U^{(0)}(\mathbf{r}_A,\mathbf{r}_B)=
 -\frac{\hbar\mu_0^2}{2\pi}\int_0^\infty
 {\mathrm d}u\,u^4\alpha_A(iu)\alpha_B(iu)
\\
\times\mathrm{Tr}\!\left[\bm{G}^{(0)}(\mathbf{r}_A,
 \mathbf{r}_B,iu)\cdot \bm{G}^{(0)}(\mathbf{r}_B,
 \mathbf{r}_A,iu)\right]
\end{multline}
is the potential observed in the case when the two atoms are in free
space, and $U^{(b)}$ is the body-induced part which can be written as
\begin{equation}
\label{us1}
U^{(b)}(\mathbf{r}_A,\mathbf{r}_B)=
 U^{(1)}(\mathbf{r}_A,\mathbf{r}_B)
 +U^{(2)}(\mathbf{r}_A,\mathbf{r}_B),
\end{equation}
where
\begin{multline}
\label{vdw1}
U^{(1)}(\mathbf{r}_A,\mathbf{r}_B)
= -\frac{\hbar\mu_0^2}{\pi}
 \int_0^\infty {\mathrm d}u\,u^4\alpha_A(iu)\alpha_B(iu)
\\
\times\mathrm{Tr}\!\left[\bm{G}^{(0)}(\mathbf{r}_A,
 \mathbf{r}_B,iu)\cdot \bm{G}^{(1)}(\mathbf{r}_B
 \mathbf{r}_A,iu)\right]
\end{multline}
is the contribution due to the cross term of the free-space part and
the scattering parts of the Green tensor, and
\begin{multline}
\label{vdw2}
U^{(2)}(\mathbf{r}_A,\mathbf{r}_B)
= -\frac{\hbar\mu_0^2}{2\pi}\int_0^\infty
 {\mathrm d}u\,u^4\alpha_A(iu)\alpha_B(iu)
\\
\times\mathrm{Tr}\!\left[\bm{G}^{(1)}(\mathbf{r}_A,
 \mathbf{r}_B,iu)\cdot \bm{G}^{(1)}(\mathbf{r}_B,
 \mathbf{r}_A,iu)\right]
\end{multline}
is the scattering-part contribution.

The free-space Green tensor reads (see, e.g., Ref.~\cite{rev})
\begin{multline}
\label{bg}
{\bm G}^{(0)}(\mathbf{r}_A ,\mathbf{r}_B,\omega)
\\
=-\frac{c^2}{4\pi
  \omega^2l^3}\left[f(-il\omega/c)\bm{I}-g(-il\omega/c)
  \frac{\mathbf{l}\mathbf{l}}{l^2}\right]e^{il\omega/c}
\end{multline}
($\mathbf{r}_A$ $\!\neq$ $\!\mathbf{r}_B$), where $\bm{I}$ denotes the
unit tensor, $\mathbf{l}$ $\!=$ $\!\mathbf{r}_B$ $\!-$
$\!\mathbf{r}_A$, $l$ $\!=$ $\!|\mathbf{l}|$, and 
\begin{align}
\label{ff}
&f(x)=1+x+x^2,\\
\label{g}
&g(x)=3+3x+x^2.
\end{align}
By substituting Eq.~(\ref{bg}) together with Eqs. (\ref{ff}) and
(\ref{g}) into Eq.~(\ref{vdw0}), the well known Casimir--Polder
interaction potential between two ground-state atoms in free space are
obtained,
\begin{multline}
\label{Bpot}
U^{(0)}(\mathbf r_A,\mathbf r_B) =
 - \frac{\hbar}{16\pi^3 \varepsilon_0^2 l^6}
 \int_0^\infty\!\! {\rm d} u \alpha_A(iu)\alpha_B(iu)
 e^{-2lu/c}
\\
\times\left[3 + \frac{6lu}{c} + \frac{5l^2u^2}{c^2}
 + \frac{2l^3u^3}{c^3}
 + \frac{l^4u^4}{c^4}
 \right].
\end{multline}
Needless to say that the scattering part of the Green tensor depends
on the specific arrangement of the bodies under consideration.


\section{Two atoms in the presence of a magneto-electric sphere}
\label{sphere}
\begin{figure}[t]
\noindent
\begin{center}
\includegraphics[width=.9\linewidth]
{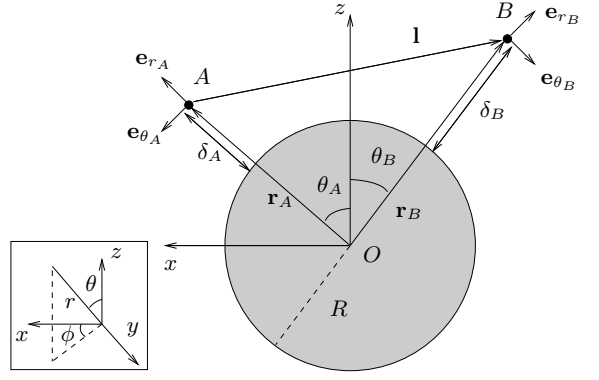}
\end{center}
\caption{
\label{figure1}
Two atoms $A$ and $B$ in the presence of a sphere
}
\end{figure}

Let us consider two atoms $A$ and $B$ in the presence of a
homogeneous magneto-electric sphere of permittivity
$\varepsilon(\omega)$, permeability $\mu(\omega)$, and radius $R$.
Choosing the coordinate system such that its origin coincides with the
center of the sphere, we may represent the scattering part of the
Green tensor as \cite{le-wei}
\begin{multline}
\label{gs}
\hspace{-1ex}
{\bm G}^{(1)}(\mathbf r_A ,\mathbf r_B,\omega)\!=\!\frac{i\omega}{4\pi
 c}\sum_{n=1}^{\infty}\frac{2n\!+\!1}{n(n\!+\!1)}
 \sum_{m=0}^n\frac{(n\!-\!m)!}{(n\!+\!m)!}(2\!-\!\delta_{0m})
\\
\times\sum_{
p=\pm 1
}\left[B_n^M(\omega)\mathbf{M}_{nm,p}
 (\mathbf{r}_A,
\omega/c
)
 \mathbf{M}_{nm,p}(\mathbf{r}_B,
\omega/c
)
\right.
\\
\left.
+B_n^N(\omega)\mathbf{N}_{nm,p}(\mathbf{r}_A,
\omega/c
)\mathbf{N}_{nm,p}(\mathbf{r}_B,
\omega/c
)\right],
\end{multline}
where $\mathbf{M}_{nm,p}(\mathbf{r},q)$ and
$\mathbf{N}_{nm,p}(\mathbf{r},q)$ are even ($p$ $\!=$ $\!+1$) and odd
($p$ $\!=$ $\!-1$) spherical wave vector functions. In spherical
coordinates, they can be expressed in terms of spherical Hankel
functions of the first kind, $h^{(1)}_n(x)$, and Legendre functions,
$P_n^m(x)$, as follows:
\begin{multline}
\label{M}
\mathbf{M}_{nm,\pm 1}(\mathbf{r},q)
=\mp\frac{m}{\sin\theta}\,
  h^{(1)}_n(qr)P_n^m(\cos\theta)
\genfrac{}{}{0pt}{}{\sin}{\cos}
(m\phi) \mathbf{e}_{\theta}
\\
-h^{(1)}_n(qr)
  \frac{\mathrm{d}P_n^m(\cos\theta)}{\mathrm{d}\theta}\,
\genfrac{}{}{0pt}{}{\cos}{\sin}
(m\phi)\mathbf{e}_\phi,
\end{multline}
\begin{multline}
\label{N}
\mathbf{N}_{nm,\pm 1}(\mathbf{r},q)=\frac{n(n+1)}{qr}\,h^{(1)}_n(qr)
  P_n^m(\cos\theta)
\genfrac{}{}{0pt}{}{\cos}{\sin}
(m\phi)\mathbf{e}_r
\\
+\frac{1}{qr}\frac{\mathrm{d}[rh^{(1)}_n(qr)]}{\mathrm{d}r}
  \left[\frac{\mathrm{d}P_n^m(\cos\theta)}{\mathrm{d}\theta}\,
\genfrac{}{}{0pt}{}{\cos}{\sin}
  (m\phi)
  \mathbf{e}_{\theta}
\right.
\\
\left.
\mp\frac{m}{\sin\theta}\,P_n^m(\cos\theta)
\genfrac{}{}{0pt}{}{\sin}{\cos}
  (m\phi)\mathbf{e}_\phi\right],
\end{multline}
with $\mathbf{e}_r$, $\mathbf{e}_\theta$, and $\mathbf{e}_\phi$ being
the mutually orthogonal unit vectors pointing in the directions of
radial distance $r$, polar angle $\theta$, and azimuthal angle $\phi$,
respectively (inset in Fig.~\ref{figure1}). The  coefficients
$B_n^M(\omega)$ and $B_n^N(\omega)$ in Eq.~(\ref{gs}) read
\begin{equation}
\label{BM}
B_n^M(\omega)=-\frac{\mu(\omega)[z_0j_n(z_0)]^\prime
  j_n(z_1)\!-\![z_1j_n(z_1)]^\prime
  j_n(z_0)}{\mu(\omega)[z_0h^{(1)}_n(z_0)]^\prime
  j_n(z_1)\!-\![z_1j_n(z_1)]^\prime
  h^{(1)}_n(z_0)}\,,
\end{equation}
\begin{equation}
\label{BN}
B_n^N(\omega)=-\frac{\varepsilon(\omega)[z_0j_n(z_0)]^\prime
  j_n(z_1)-[z_1j_n(z_1)]^\prime j_n(z_0)}{\varepsilon(\omega)
  [z_0h^{(1)}_n(z_0)]^\prime j_n(z_1)-[z_1j_n(z_1)]^\prime
  h^{(1)}_n(z_0)},
\end{equation}
where $j_n(z)$ is the spherical Bessel function of the first kind,
$z_0$ $\!=$ $\!kR$ ($k$ $\!=$ $\!\omega/c$),
$z_1$ $\!=$ $\!\sqrt{\varepsilon(\omega)\mu(\omega)}z_0$
$\!=$ $\!n(\omega)z_0$, and the prime denotes differentiation with
respect to the respective argument.
Without loss of generality, we assume that the two atoms are located
in the $xz$-plane (Fig.~\ref{figure1}),
\begin{equation}
\label{position}
\mathbf{r}_A=(r_A,\theta_A,0),\quad \mathbf{r}_B=(r_B,\theta_B,\pi).
\end{equation}

Substituting  Eqs.~(\ref{M}) and (\ref{N}) [together with
Eq.~(\ref{position})] into Eq.~(\ref{gs}), and performing the
summations over $m$ and $p$, we derive (Appendix~\ref{grn-sc})
\begin{equation}
\label{23}
{\bm G}^{(1)}(\mathbf{r}_A,\mathbf{r}_B,iu)=
\sum_{ij}
 G_{ij}^{(1)}(\mathbf{r}_A,\mathbf{r}_B,iu)
 \mathbf{e}_{i_A}\mathbf{e}_{j_B}
\end{equation}
($i,j$ $\!=$ $\!r,\theta,\phi$), with the nonzero elements being
\begin{multline}
\label{G1}
G^{(1)}_{rr}(\mathbf{r}_A,\mathbf{r}_B,\omega)
= \frac{ic}
 {4\pi \omega r_A
 r_B}
\\
\times
\sum_{n=1}^\infty n(n\!+\!1)(2n\!+\!1)B_n^N(\omega)
 P_n(\gamma)Q_n^{(1)},
\end{multline}
\begin{multline}
\label{G2}
G^{(1)}_{
\genfrac{}{}{0pt}{}{r\theta}{\theta r}
}
(\mathbf{r}_A,\mathbf{r}_B,\omega)
\\
=\frac{-ic\sin\Theta}
 {4\pi \omega r_A r_B}\sum_{n=1}^\infty(2n\!+\!1)
 B_n^N(\omega)P_n'(\gamma)
\genfrac{}{}{0pt}{}{Q_n^{(2)}}{Q_n^{(3)}},
\end{multline}
\begin{multline}
\label{G4}
G^{(1)}_{
\genfrac{}{}{0pt}{}{\theta\theta}{\phi\phi}
}
(\mathbf{r}_A,\mathbf{r}_B,\omega)
 =-\frac{i\omega}{4\pi  c}
 \sum_{n=1}^\infty\frac{2n+1}{n(n+1)}
\\
\times\left[B_n^M(\omega)
\genfrac{}{}{0pt}{}{P'_n(\gamma)}{F_n(\gamma)}
Q_n^{(1)}+\frac{c^2B_n^N(\omega)}
 {\omega^2 r_A r_B}
\genfrac{}{}{0pt}{}{F_n(\gamma)}{P_n'(\gamma)}
Q_n^{(4)}\right],
\end{multline}
where $\Theta$ $\!=$ $\!\theta_A$ $\!+$ $\!\theta_B$,
$\gamma$ $\!=$ $\!\cos\Theta$, and
\begin{equation}
\label{Q1}
 Q^{(1)}_n=h_n^{(1)}(kr_A)h_n^{(1)}(kr_B),
\end{equation}
\begin{equation}
\label{Q2}
Q_n^{(2)}=h_n^{(1)}(kr_A)[zh_n^{(1)}(z)]'_{z=kr_B},
\end{equation}
\begin{equation}
\label{Q3}
Q_n^{(3)}=h_n^{(1)}(kr_B)[yh_n^{(1)}(y)]'_{y=kr_A},
\end{equation}
\begin{equation}
\label{Q4}
Q_n^{(4)}=[yh_n^{(1)}(y)]'_{y=kr_A}[zh_n^{(1)}(z)]'_{z=kr_B},
\end{equation}
\begin{equation}
\label{FFF}
F_n(x)=n(n+1) P_n(x) -x P_n'(x).
\end{equation}
To facilitate further evaluations, it is convenient to represent
the free-space Green tensor~(\ref{bg}) in the same spherical
coordinate system as the scattering part, so that its nonzero elements
read
\begin{equation}
\label{G6}
G_{rr}^{(0)}(\mathbf{r}_A,\mathbf{r}_B,iu)=
 \frac{c^2}{4\pi u^2l^5}\big[l^2f(\xi)\cos\Theta-g(\xi)
 l_Al_B\big]e^{-\xi},
\end{equation}
\begin{multline}
\label{G7}
G_{r \theta(\theta r)}^{(0)}
(\mathbf{r}_A,\mathbf{r}_B,iu)
=\frac{-c^2\sin\Theta }{4\pi u^2l^5}
\\
\times\left[l^2f(\xi)\pm g(\xi)
 r_{A(B)}l_{A(B)}\right]e^{-\xi},
\end{multline}
\begin{multline}
\label{G7.5}
G_{\theta\theta }^{(0)}
(\mathbf{r}_A,\mathbf{r}_B,iu)
=\frac{-c^2}{4\pi u^2l^5}
\\
\times\left[l^2f(\xi)\cos\Theta -
g(\xi)r_Ar_B\sin^2\Theta\right]e^{-\xi},
\end{multline}
\begin{equation}
\label{G8}
G_{\phi\phi}^{(0)}(\mathbf{r}_A,\mathbf{r}_B,iu)=
 -\frac{c^2}{4\pi u^2l^3}\,f(\xi)e^{-\xi},
\end{equation}
where $\xi$ $\!=$ $\!lu/c$, $l_A$ $\!=$ $\!r_B \cos \Theta$ $\!-$
$\!r_A$ and $l_B$ $\!=$
$\!r_B$ $\!-r_A \cos\Theta$.

Recalling Eqs.~(\ref{us1})--(\ref{vdw2}), we may write the
body-induced part of the interaction potential as
\begin{equation}
\label{us2}
U^{(b)}(\mathbf{r}_A,\mathbf{r}_B)=
 \sum_{i,j}
\left[
 U^{(
1
)}_{ij}(\mathbf{r}_A,\mathbf{r}_B)
+  U^{(2)}_{ij}(\mathbf{r}_A,\mathbf{r}_B)
\right],
\end{equation}
where
\begin{multline}
\label{vdw12}
U^{(1)}_{ij}(\mathbf{r}_A,\mathbf{r}_B)
=-\frac{\hbar\mu_0^2}{\pi}
  \int_0^\infty \mathrm{d}u\,u^4\alpha_A(iu)\alpha_B(iu)
\\
\times G^{(0)}_{ij}(\mathbf r_A,\mathbf
  r_B,iu)G^{(1)}_{ij}(\mathbf{r}_A,\mathbf{r}_B,iu),
\end{multline}
\begin{multline}
\label{vdw22}
U^{(2)}_{ij}(\mathbf{r}_A,\mathbf{r}_B)
 =-\frac{\hbar\mu_0^2}{2\pi}
 \int_0^\infty \mathrm{d}u\,u^4\alpha_A(iu)\alpha_B(iu)
\\
\times\left[G^{(1)}_{ij}(\mathbf{r}_A,
  \mathbf{r}_B,iu)\right]^2,
\end{multline}
with $G^{(1)}_{ij}(\mathbf{r}_A,\mathbf{r}_B,iu)$ and
$G^{(0)}_{ij}(\mathbf{r}_A,\mathbf{r}_B,iu)$ according to
Eqs.~(\ref{G1})--(\ref{G4}) and Eqs.~(\ref{G6})--(\ref{G8}),
respectively (see Appendix~\ref{Uij}). Further evaluation of
$U^{(b)}(\mathrm{r}_A,\mathrm{r}_B)$ requires numerical methods in
general. Before doing so, let us consider the limiting cases of large
and small spheres.


\subsection{Large sphere}
\label{secA}

The limiting case of a large sphere may be defined by the requirement
that
\begin{equation}
\label{dr}
\delta_{A'}\equiv r_{A'}-R\ll R
\quad
(A' = A,B)
\end{equation}
and
\begin{equation}
\label{theta}
l \ll R
\quad\leadsto\quad
\Theta = \theta_A + \theta_B \ll 1
\end{equation}
(cf. Fig.~\ref{figure1}). In this limit, Eq.~(\ref{us2}) leads to
(Appendix~\ref{ls})
\begin{multline}
\label{l-e}
U^{(b)}(\mathbf{r}_A,\mathbf{r}_B)=
\\
\frac{\hbar}{16 \pi^3
 \varepsilon_0^2l^5l_+^5}
 \biggl\{\left[4X^4-2\delta_-^2\delta_+^2+X^2(\delta_-^2
 +\delta_+^2)\right]I_{01}
\\
+ \frac{l_+^2}{4R}\left[3(l_+^3-\delta_+^3)-\delta_+
 (\delta_-^2+4X^2)\right]I_{11}
\\
-\frac{3l^5}{l_+}\left(I_{02}
+\frac{l_+}{4R}I_{12}\right)\biggr\},
\end{multline}
where $X$ $\!=$ $\! -R\Theta$, $\delta_{\pm}$ $\!=$ $\!\delta_B$ $\!\pm$
$\!\delta_A$, $l_+$ $\!=$ $\!(X^2$ $\!+$ $\!\delta_{+}^2)^{1/2}$, and
\begin{multline}
I_{kl}=
\int_0^\infty \mathrm{d} u\, \alpha_A(iu)\alpha_B(iu)
\\\times
\left[26+\varepsilon(iu)\mu(iu)R^2u^2/c^2\right]^k
\left[\frac{\varepsilon(iu)-1}{\varepsilon(iu)+1}\right]^l.
\end{multline}
Note that the case of a purely electrically polarizable sphere can be
simply obtained by setting $\mu(iu)$ $\!=$ $\!1$ in Eq.~(\ref{l-e}).

For a purely magnetically polarizable sphere \mbox{[$\varepsilon(iu)$
$\!=$ $\!1$]} Eq.~(\ref{us2}) leads, under the conditions
(\ref{dr}) and (\ref{theta}), to (Appendix~\ref{ls})
\begin{multline}
\label{l-m}
U^{(b)}(\mathbf{r}_A,\mathbf{r}_B)=\frac{\hbar\left[\delta_-^2-2
 X^2+3\delta_+(l_+ -\delta_+)\right]}{64 \pi^3
 \varepsilon_0^2c^2l^5l_+}
\\
\times\int_0^\infty \mathrm{d} u\,u^{2}\alpha_A(iu)
 \alpha_B(iu) \frac{\left[\mu(iu)-1][\mu(iu)-3\right]}
 {\mu(iu)+1}.
\end{multline}
As expected, Eqs.~(\ref{l-e}) and (\ref{l-m}) are in agreement with
those found for a half-space ($l_+/R$ $\!\rightarrow$ $\!0$)
\cite{h}.


\subsection{Small spheres}

In the opposite limit of a small sphere, where
\begin{equation}
\label{smalls}
R\ll r_{A'}
\quad
(A' = A,B),
\end{equation}
Eq.~(\ref{us2}) leads to (Appendix~\ref{ls})
\begin{multline}
\label{s}
U^{(b)}(\mathbf r_A, \mathbf r_B)
\\
=\frac{\hbar}
 {64\pi^4\varepsilon_0^3r_A^3r_B^3l^3}
 \int_0^\infty
 \mathrm{d}u\alpha_A(iu)\alpha_B(iu)e^{-(r_A+r_B+l)u/c}
\\
\times\!\biggl\{\alpha_{\mathrm{sp}}(iu)
 \biggl[f(\xi)\big\{g(b)\big[2(1\!+\!a)\!
 \!-\!g(a)\sin^2\Theta\big]\!+\!2a^2f(b)\big\}
\\
+\frac{g(\xi)}{l^2}
 \big\{\big[(2l^2\!-\!r_A r_B
 \cos\Theta)f(a)f(b)\!+\!2a^2f(b)r_Al_A
\\
 -\!2b^2f(a)r_Bl_B\big]\sin^2\Theta\!-\!4(1\!+\!a)(1\!+\!b)l_A
 l_B\cos\Theta \big\}\biggr]
\\
 +\frac{a b}{c^2}(1\!+\!a)(1\!+\!b)
 \beta_{\mathrm{sp}}(iu)
\biggl[g(\xi)\frac{r_Ar_B}{l^2}\sin^2\Theta
\\
 -2f(\xi)
\cos\Theta\biggr]\biggr\}
\end{multline}
($a$ $\!=$ $\!r_Au/c$, $b$ $\!=$ $\!r_Bu/c$),
where
\begin{align}
\label{a-s}
& \alpha_{\mathrm{sp}}(\omega)=4\pi
 \varepsilon_0R^3\,\frac{\varepsilon(\omega)-1}{\varepsilon(\omega)+2},
\\[1ex]
\label{b-s}
& \beta_\mathrm{sp}(\omega)=\frac{4\pi  R^3}{\mu_0}\,
 \frac{\mu(\omega)-1}{\mu(\omega)+2}.
\end{align}
Let us consider a sphere to which the Clausius--Mossotti relation
applies, so that
\begin{equation}
 \frac{\varepsilon(\omega)-1}{\varepsilon(\omega)+2}
 =\frac{1}{3\varepsilon_0} \sum_k n_k \alpha_k(\omega),
 \end{equation}
with $n_k$ and $\alpha_{k}(\omega)$, respectively, being
the number density and the polarizability of the atoms
of type $k$. In this case, Eq.~(\ref{a-s}) can be rewritten as
\begin{equation}
\label{e48}
\alpha_{sp}(\omega)= \sum_{k} n_k \alpha_k(\omega),
\end{equation}
where $N_k$ is the number of atoms of type $k$ of the sphere.
Accordingly, the magnetic analog of Eq.~(\ref{e48}) is
\begin{equation}
\label{e49}
\beta_{sp}(\omega)= \sum_{k} n_k \beta_k(\omega),
\end{equation}
with $\beta_k(\omega)$ being the magnetizability of the atoms
of type~$k$. Hence, we may replace in Eq.~(\ref{s}) the sphere
parameters $\alpha_\mathrm{sp}(iu)$ and $\beta_\mathrm{sp}(iu)$,
respectively, with the electric and magnetic polarizability of a
single atom [say $\alpha_k(iu)$ and $\beta_k(iu)$], to obtain the
nonadditive interaction potential of three atoms, two of which being
purely electrically polarizable whereas the third atom being
simultaneously electrically and magnetically polarizable. Indeed,
after a straightforward but lengthy calculation, it can be shown that
in the case of a purely electrically polarizable sphere, Eq.~(\ref{s})
[$\beta_\mathrm{sp}(iu)$ $\!=$ $\!0$] leads to the interaction
potential between three electrically polarizable atoms,
as derived in Refs.~\cite{Ax,Pow1,Pow2,Riz}.

In the retarded limit where $l$,$r_A$,$r_B$ $\!\gg$
$\!c/\omega_{\mathrm{min}}$ ($\omega_{\mathrm{min}}$ denoting the
minimum frequency among the relevant atomic and medium transition
frequencies), due to the presence of the exponential term in the
integral in Eq.~(\ref{s}), only small values of $u$ significantly
contribute to the integral. Therefore, the electric and magnetic
polarizabilities can be approximately replaced with their static
values. After performing the remaining integral and expressing all the
geometric parameters in terms of $r_A$, $r_B$, and $l$, we arrive at
\begin{multline}
\label{s-ret}
U^{(b)}(\mathbf r_A, \mathbf r_B)=\frac{\hbar c
 \alpha_A(0)\alpha_B(0)}{32\pi^4\varepsilon_0^3r_A^5r_B^5l^5
 (r_A+r_B+l)^7}
\\
\times\biggl[\alpha_\mathrm{sp}(0)
 \bigl\{
 \mathcal{S}[h_1(r_A,r_B,l)]
 +\mathcal{S}[h_1(r_B,l,r_A)]
\\
+\mathcal{S}[h_1(l,r_A,r_B) ]\bigr\}
 +\frac{r_A^2r_B^2}{c^2}\beta_\mathrm{sp}(0)
 \mathcal{S}[h_2(r_A,r_B,l)]\biggr],
\end{multline}
where
\begin{multline}
h_1(x,y,z)=3x^6y^2(y\!-\!x)(x\!+\!y\!+\!7z)(x^2\!+\!7xy\!
 +\!11y^2)
\\
- x^4y^2z^2(53x^4 + 280x^3y - 137x^2y^2
\\
 -329xy^3 - 623xy^2z - 192y^2z^2),
\end{multline}
\begin{multline}
h_2(x,y,z)=3x^4(y\!-\!x)(x\!+\!y\!+\!7z)(x^2\!+\!7xy\!+\!11y^2)
\\
-\!2x^3z^2(x\!+\!y)(26x^2\!+\!93xy\!-\!133y^2)
 \!-\!7x^2z^5(3x\!-\!2y)
\\
-\!14x^3z^3(2x^2\!-\!3xy\!-\!13y^2)\!-\!x^3z^4(17x\!+\!161y)
\\
+\!2xz^6(31x\!+\!105y)\!+\!5z^7(14x\!+\!z),
\end{multline}
and $\mathcal{S}[f(x,y,z)]$ $\!=$ $f(x,y,z)$ $\!+$ $\!f(y,x,z)$.

\begin{figure}
\noindent

\begin{center}
\includegraphics[width=0.5\linewidth]{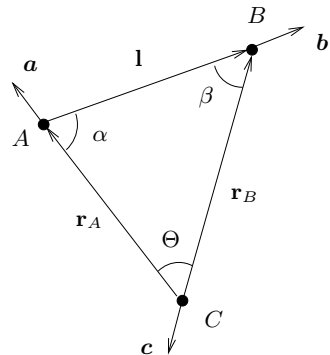}
\end{center}
\caption{
\label{triangle}
The triangle formed by the two atoms (at corners $A$ and $B$)
and the sphere (at the corner $C$) in the small-sphere limit.
It is seen that the vector products ${\bm{a}}\!\cdot\! {\bm{b}}$,
${\bm{b}}\!\cdot\!{\bm{c}}$ and ${\bm{c}}\!\cdot\! {\bm{a}}$ in the
Axilrod and Teller's formula \cite{Ax} are equal to $-\cos\alpha$,
$-\cos\beta$, and $-\cos\Theta$, respectively.
}
\end{figure}

In the nonretarded limit where $l$,$r_A$,$r_B$ $\!\ll$
$\!c/[n(0)\omega_{\mathrm{max}}]$ ($\omega_{\mathrm{max}}$ denoting
the maximum frequency among the relevant atomic and medium transition
frequencies), the leading contribution to the integral in
Eq.~(\ref{s}) comes from the region where $e^{-(r_A+r_B+l) u/c}$
$\!\simeq$ $\!1$, so Eq.~(\ref{s}) reduces to
\begin{multline}
\label{s-nret}
U^{(b)}(\mathbf r_A, \mathbf r_B)
=\frac{3\hbar}{64\pi^4\varepsilon_0^3r_A^3r_B^3l^3}
\\
\times\left\{\left[1\!-\!\frac{1}{l^2}(4l_Al_B\!+\!
 r_Ar_B\sin^2\Theta)\cos\Theta+\!\cos^2\Theta\!\right]
J
_1
\right.
\\
\left.
+\frac{r_Ar_B}{c^4}
 \left(\frac{r_Ar_B}{l^2}\sin^2\Theta\!-\!\frac{2}{3}
 \cos\Theta\right)
J
_2\right\},
\end{multline}
where
\begin{align}
\label{I1}
&
J_1=\int_0^\infty
 \mathrm{d}u\,\alpha_A(iu)\alpha_B(iu)\alpha_\mathrm{sp}(iu),
\\[1ex]
\label{I2}
&
J
_2=\int_0^\infty
 \mathrm{d}u\,u^2\alpha_A(iu)\alpha_B(iu)\beta_\mathrm{sp}(iu).
\end{align}
In particular, in the case of a purely electrically polarizable sphere
($J_2$ $\!=$ $\!0$), Eq.~(\ref{s-nret}) can be written in a very
symmetric form. For this purpose we introduce the unit vectors
${\bm{a}}$, ${\bm{b}}$, and ${\bm{c}}$ pointing in the directions of
$\mathbf{r}_A$, $\mathbf{l}$, and $-\mathbf{r}_B$, respectively
(see Fig.~\ref{triangle}). Noting that $l_A$ and $l_B$ defined below
Eq.~(\ref{G8}) are the components of the vector $\mathbf{l}$ in the
directions of $\mathbf{r}_A$ and $\mathbf{r}_B$ and can thus be
written as $l({\bm{a}}\!\cdot\!{\bm{b}})$ and
$-l({\bm{b}}\!\cdot\! {\bm{c}})$, respectively, we see that
\begin{equation}
\label{d2}
-\frac{1}{l^2}(4l_Al_B\!+\!
 r_Ar_B\sin^2\Theta)+\!\cos\Theta\!=\!
 3({\bm{a}}\!\cdot\! {\bm{b}})({\bm{b}}\!\cdot\! {\bm{c}})
\end{equation}
and can rewrite Eq.~(\ref{s-nret}) as
\begin{equation}
\label{d1}
U^{(b)}(\mathbf r_A, \mathbf r_B)
=\frac{3\hbar}{64\pi^4\varepsilon_0^3r_A^3r_B^3l^3}
 \left[1-3({\bm{a}}\!\cdot\! {\bm{b}})
  ({\bm{b}}\!\cdot\!{\bm{c}})
 ({\bm{c}}\!\cdot\! {\bm{a}})\right]
 J_1.
\end{equation}
If $\alpha_{sp}(iu)$ in Eq.~(\ref{I1}) is again identified with the
the electric polarizability of a single atom, Eq.~(\ref{d1}) is
nothing but the formula for the nonretarded three-atom interaction
potential, which was first given by Axilrod and Teller \cite{Ax}.
%

\section{Numerical results}
\label{num}

\begin{figure}[t]
\noindent
\begin{center}
\includegraphics[width=.9\linewidth]
{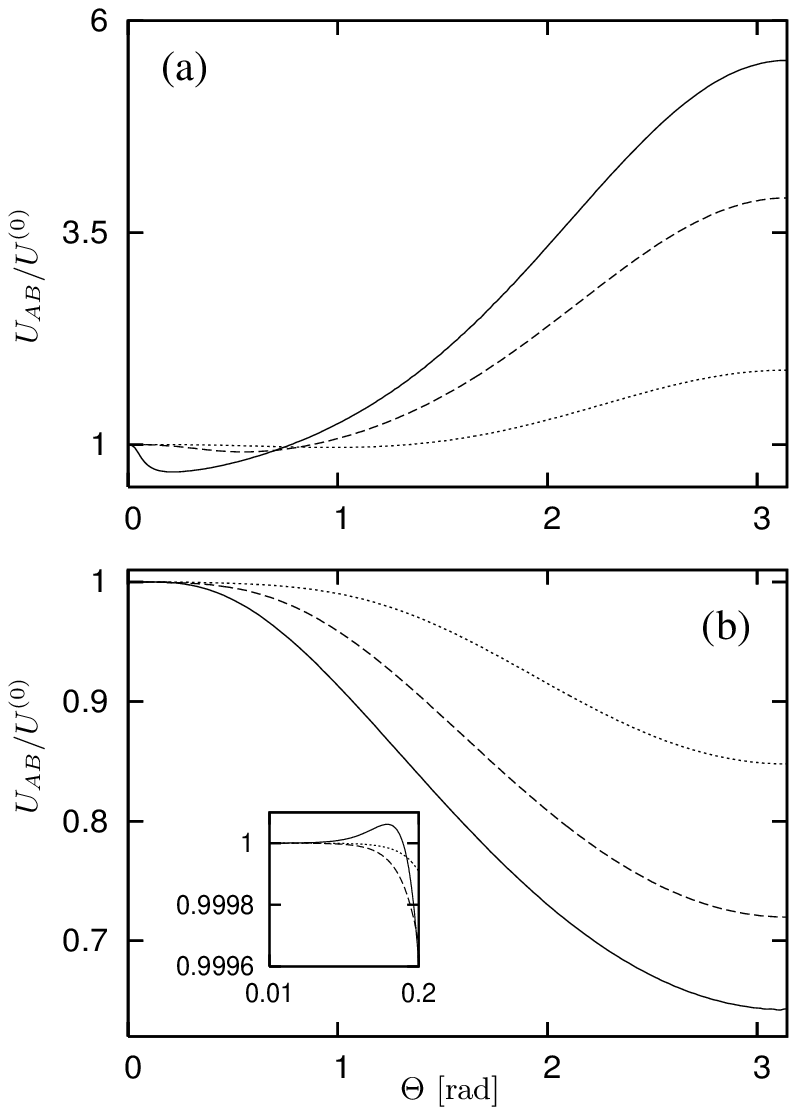}
\end{center}
\caption{
\label{tri}
The mutual vdW potential of two identical two-level atoms in a
triangular configuration with (a) a purely electrically polarizable
sphere with
\mbox{$\omega_{\mathrm{P}e}/\omega_{10}$ $\!=$ $\!3$},
\mbox{$\omega_{\mathrm{T}e}/\omega_{10}$ $\!=$ $\!1$},
and \mbox{$\gamma_e/\omega_{10}$ $\!=$ $\!0.001$}
and (b) a purely magnetically polarizable sphere with
\mbox{$\omega_{\mathrm{P}m}/\omega_{10}$ $\!=$ $\!3$},
\mbox{$\omega_{\mathrm{T}m}/\omega_{10}$ $\!=$ $\!1$},
and \mbox{$\gamma_m/\omega_{10}=0.001$} is shown as a function of the
atom--atom angular separation $\Theta$ ($\omega_{10}$ is the atomic
transition frequency). The sphere radius is $R$ $\!=$
$\!c/\omega_{10}$ and the distances between the atoms and the center
of the sphere are $r_{A}$ $\!=$ $\!r_{B}$ $\!=$
$\!1.03\,c/\omega_{10}$ (solid line), $1.3\,c/\omega_{10}$ (dashed
line), and $2\,c/\omega_{10}$ (dotted line). $U^{(0)}$ is the
potential observed in free space.
}
\end{figure}

\begin{figure}
\noindent
\begin{center}
\includegraphics[width=1.\linewidth]
{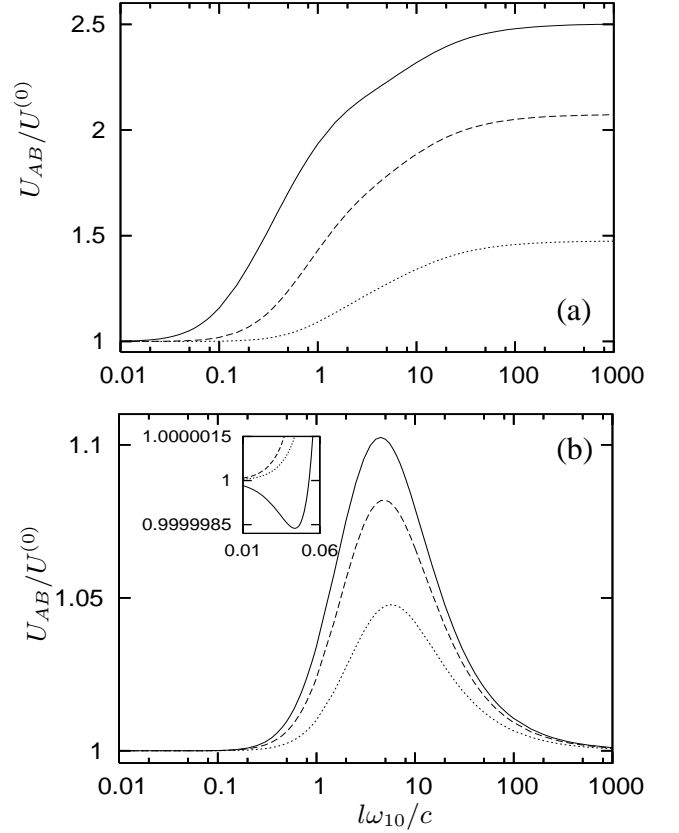}
\end{center}
\caption{
\label{lin}
The mutual vdW potential of two identical two-level atoms in a linear
configuration with (a) a purely electrically polarizable sphere and
(b) a purely magnetically polarizable sphere is shown as a function of
the interatomic distance $l$. Atom $A$ is held at a fixed position
between atom $B$ and the sphere center with $r_{A}\!$ $\!=$
$\!1.03\,c/\omega_{10}$ (solid line), $1.1\,c/\omega_{10}$ (dashed
line), and $1.3\,c/\omega_{10}$ (dotted line). All other parameters
are the same as in Fig.~\ref{tri}.
}
\end{figure}

The effect of a medium-sized magneto-electric sphere on the mutual vdW
interaction of two identical two-level atoms is illustrated in
Figs.~\ref{tri} and \ref{lin} showing the ratio $U_{AB}$ $/U^{(0)}$
[cf.~Eq.~(\ref{uparts})]. The results have been found by exact
numerical evaluation of Eq.~(\ref{us2}) together with
Eqs.~(\ref{u1rr})--(\ref{u2tt}), where the permittivity and
permeability of the sphere have been described by single-resonance
Drude--Lorentz models,
\begin{align}
\label{epsilon}
&\varepsilon(\omega)=1+\frac{\omega_{{\rm P}e}}
         {\omega_{{\rm T}e}^2-\omega^2
 -i\gamma_e \omega},
\\
\label{mu}
&\mu(\omega)=1+\frac{\omega_{{\rm P}m}}{\omega_{{\rm T}m}^2-\omega^2
 -i\gamma_m \omega}.
\end{align}

In Fig.~\ref{tri}, a configuration is considered where the two atoms
are positioned at equal distances from the sphere, $r_A$ $\!=$
$\!r_B$, briefly referred to as triangular configuration, and
$U_{AB}/U^{(0)}$ is shown as a function of the angular separation
$\Theta$ for three different values of the atom--sphere separation.
For a purely electrically polarizable sphere [Fig.~\ref{tri}(a)],
depending on the separation angle between the atoms, a (compared
to the free-space case) relative reduction or enhancement of
the vdW potential is possible, while for a purely magnetically
polarizable sphere [Fig.~\ref{tri}(b)], the potential is typically
reduced [note that for very small angular separations, a slight
enhancement is possible, as can be seen from the inset in
Fig.~\ref{tri}(b)], and the reduction increases with the angular
separation. In both cases, the sphere-induced modification is
strongest when the atoms are at opposite sides of the sphere
($\Theta$ $\!=$ $\!\pi$). Note that for small atom--sphere separations
(solid curves) and small angular separations, the potential
qualitatively agrees with the potential obtained for two atoms placed
in parallel alignment near a semi-infinite half space \cite{h}, as
expected from the results in Sec.~\ref{secA}.

In Fig.~\ref{lin}, a configuration is considered where the two atoms
and the sphere center are aligned on a straight line, briefly referred
to as linear configuration, and $U_{AB}/U^{(0)}$ is shown as a
function of the interatomic distance for three different values of
the position $r_A$ of atom $A$ which is positioned between the sphere
and atom $B$. Unless both atoms are very close to the sphere, the
sphere gives always rise to a (compared to the free-space case)
relative enhancement of the vdW potential between the atoms; only for
very small atom--sphere separations the potential can be reduced if
the sphere is purely magnetically polarizable [inset in
Fig.~\ref{lin}(b)]. Figure~\ref{lin}(a) shows that in the presence of
a purely electrically polarizable sphere the relative enhancement
of the potential increases with the interatomic separation $l$ and
approaches a limit for larger interatomic separations, which depends
on the separation distance between atom $A$ and the sphere. From
Fig.~\ref{lin}(b) it is seen that in the presence of a purely
magnetically polarizable sphere the relative enhancement of the
potential increases with the interatomic separation $l$, reaches a
maximum, and decreases with a further increase of $l$. In agreement
with the results of Sec.~\ref{secA}, the potential observed for small
atom sphere separations (solid curves) and small interatomic
separations qualitatively agrees with the potential obtained for two
atoms placed in vertical alignment near a semi-infinite half space
\cite{h}.

Many features of the vdW potential observed in Figs.~\ref{tri} and \ref{lin}
 can be subject to a physical interpretation via the method of image
charges (the same approach as has been used for a planar geometry Ref.
\cite{h}).
Although being strictly valid only for sufficiently small atom--atom
and atom--surface distances (such that retardation is negligible) and
being most easily applicable in the perfect conductor limit, this
approach yields qualititive predictions for the sphere-induced
enhancement and reduction of the potential which apply beyond this
case. 
According to the image-charge method,
the effect of the boundaries is simulated by suitably placed image
charges of appropriate magnitudes, so that the two-atom vdW potential
effectively consists of interactions between fluctuating dipoles $A$
and $B$ and their images ${A}'$ and ${B}'$ in the sphere, with
\begin{equation}
 \hat H_{\mathrm {int}}=\hat V_{AB}+\hat V_{AB'}+\hat
V_{BA'}
\end{equation}
being the corresponding interaction Hamiltonian. Here, $\hat{V}_{AB}$
denotes the direct interaction between dipole $A$ and dipole $B$,
while $\hat V_{AB'}$ and $\hat V_{BA'}$ denote the indirect
interaction between each dipole and the image induced by the other one
in the sphere. The leading contribution to the energy shift is of
second order in $\hat{H}_{\mathrm {int}}$,
\begin{multline}
\label{dE}
\Delta E_{AB}=-\sideset{}{'}\sum_{n,m}
\frac{\langle 0_{A},0_{B}|\hat H_{\mathrm {int}}
|n_{A},m_{B}\rangle}{\hbar(\omega_{A}^n+
\omega_{B}^m)}
\\
\times\,\langle n_{A},
m_{B}|\hat H_{\mathrm {int}}|0_{A},
0_{B}\rangle,
\end{multline}
where $|n_{A(B)}\rangle$ denotes the energy eigenstates of atom $A(B)$
with eigenenergies $\omega_{A(B)}^n$ and the prime indicates that the
terms $n_{A(B)}=0_{A(B)}$ are not included in the sum.

In this approach, $U^{(0)}$ corresponds to the product of two
direct interactions and is negative in accordance with
Eq.~(\ref{dE}). $U^{(2)}$ is the product of two indirect interactions
and is also negative. The terms containing one direct and one indirect
interaction are contained in $U^{(1)}$. Since the total potential is
equal to $U_{AB}=U^{(0)}+U^{(1)}+U^{(2)}$, the 
sum $U^{(1)}+U^{(2)}$ represents the effects of the medium. The 
relative signs and strengths of $U^{(1)}$ and $U^{(2)}$ will 
determine
whether the free space vdW interaction is enhanced or suppressed.
\begin{figure}
\noindent
\begin{center}
\includegraphics[width=.9\linewidth]
{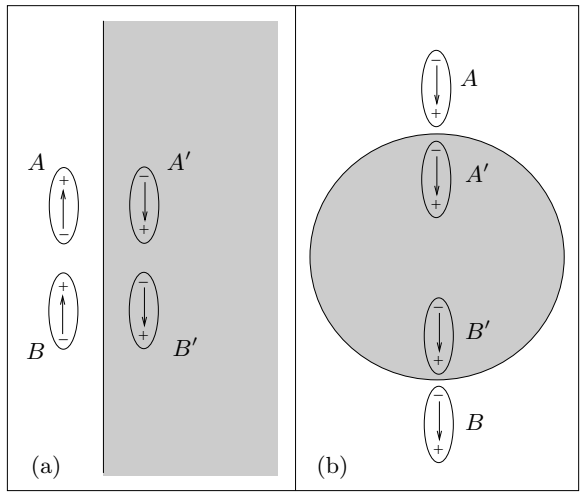}
\end{center}
\caption{\label{f5}
Two electric dipoles near a purely electrically
polarizable sphere are shown in a triangulare configuration where the
dipole-dipole angular separation (a) is
small enough to replace the
sphere, approximately, by a halfspace and (b) is equal to $\pi$.
}
\end{figure}
The orientations of the dipoles $A$ and $B$ are random and
independent of each other, so that strictly speaking the signs of all
dipole--dipole interactions has to be obtained by averaging over all
possible orientations. The effect of such averaging on the sign of the
interactions can be reproduced by restricting the attention to the
maximally attractive case of both dipoles pointing in the same
direction parallel to their connecting line, with the dipole--dipole
interaction $\hat{V}_{AB}$ being negative in this case. The image
dipoles ${A}'$ and ${B}'$ are constructed by appropriate reflection of
the dipoles $A$ and $B$. The resulting signs of the interactions $\hat
V_{AB'}$ and $\hat V_{BA'}$ between dipoles and image dipoles are
negative/positive if the respective dipole moments are
parallel/antiparallel. 
\begin{figure}
\noindent
\begin{center}
\includegraphics[width=.9\linewidth]
{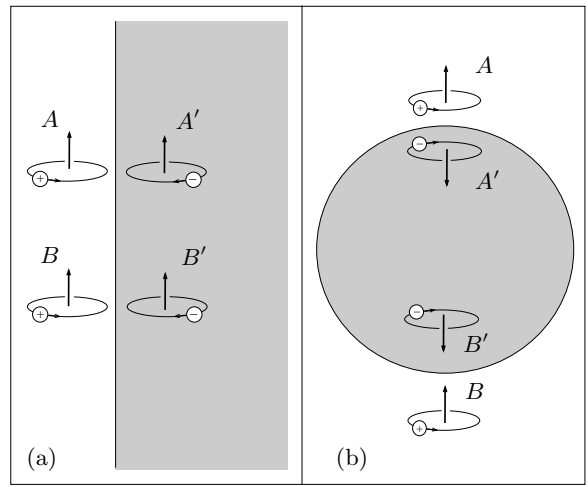}
\end{center}
\caption{\label{f6}
Two magnetic dipoles near a purely electrically
polarizable sphere are shown in a triangulare configuration
for a (a) very small angular separation (b) maximal angular separation.
}
\end{figure}
Figure~\ref{f5} shows two electric dipoles near a purely electrically
polarizable sphere in the triangular configuration together with
their images in the sphere. When the inter-dipole angular separation
is very small, the curvature of the spherical surface can be
disregarded and the sphere can be approximately replaced by a half
space as in Fig.~\ref{f5}(a). It is seen that $U^{(1)}$, which is a product
of one indirect and one direct interaction, is positive. Since the
negative $U^{(2)}$ is a product of two indirect interactions, and the
direct interaction is stronger than indirect one
for small interatomic separations
the sum $U^{(1)}+U^{(2)}$ is positive 
and hence
the total potential $U_{AB}$ is weakened compared to that in free
space. This confirms the numerical results for short distances
presented in Fig.~\ref{tri}(a). The case of two dipoles located at the
opposite ends of a sphere diameter is sketched in Fig.~\ref{f5}(b). It
can be seen that $U^{(1)}$ is negative. As a consequence, $U_{AB}$ is
enhanced in agreement with the curves in Fig.~\ref{tri}(a).

The case of two electric dipoles near a purely magnetically 
polarizable sphere can be treated by considering two magnetic dipoles
near a purely electrically polarizable sphere, since the two
situations  are equivalent due to the duality between electric
and magnetic fields. 
Again we consider 
the triangular configuration. In the 
limit
of small separation angles,
the
surface can be regarded as flat. The sketch of the dipoles and 
their images in Fig.~\ref{f6}(a) indicates a negativeness of $U^{(1)}$
leading to an enhancement of the overall interaction potential
[cf. Fig.~\ref{tri}(b) inset, solid curve]. When the separation angle is
large (atoms located on opposite sides of the sphere),
it can be inferred from Fig.~\ref{f6}(b) that $U^{(1)}$ is
positive, resulting in a reduction of the interaction potential, in
agreement with the numerical results shown in Fig.~\ref{tri}(b). 

\begin{figure}
\noindent
\begin{center}
\includegraphics[width=.9\linewidth]
{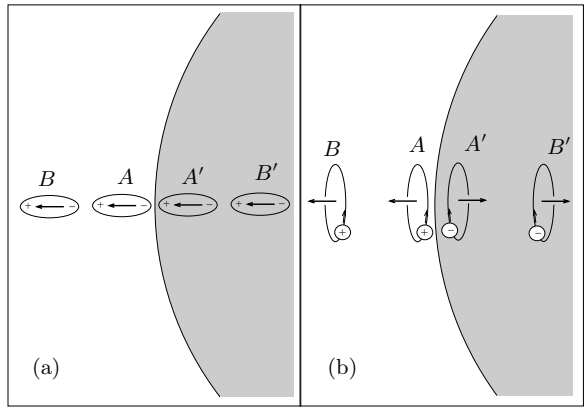}
\end{center}
\caption{\label{f7}
Two dipoles near a sphere are shown in a linear configuration.
Figure~\ref{f7}(a) is showing two electric
dipoles while in Fig.~\ref{f7}(b) two magnetic dipoles are shown.  
}
\end{figure}

We turn now to the linear configuration. For dipoles situated near a
purely electrically polarizable sphere [Fig.~\ref{f7}(a)], $U^{(1)}$ is 
negative resulting in an enhancement of the total interaction 
potential for all distance regimes
as visible in Fig.~\ref{lin}(a). For a
purely magnetically polarizable
sphere, we again 
invoke the duality principle
to replace it by a purely electrically polarizable one, and the
electric dipoles by magnetic ones as shown in Fig.~\ref{f7}(b). It can be
inferred from the  sketch that $U^{(1)}$ is positive for all
distances.  In
order to be conclusive about the body-induced effects, one 
hence
has to compare the magnitudes of the competing $U^{(1)}$ and
$U^{(2)}$.
For small atom--atom separations, the direct interaction dominates, so
$U^{(1)}$ is stronger than $U^{(2)}$ and 
the potential is reduced as shown in Fig.~\ref{lin}(b), inset. As the 
interatomic separation increases,
the indirect interaction gains in relevance and hence
$U^{(2)}$ may become dominant leading to an enhancement of the
total vdW potential, in agreement with the curves presented in
Fig.~\ref{lin}(b).

\section{Summary}
\label{summ}

We have studied the mutual vdW interaction between two atoms near a
dispersing and absorbing magneto-electric sphere and presented both
analytical and numerical results. When the radius of the sphere
becomes sufficiently large, then the interaction potential tends to
the one found for two atoms near a magneto-electric half space. In the
opposite case of a very small sphere, the sphere can be regarded as
being a third atom with respective electric and magnetic
polarizabilities. In particular for electrically polarizable atoms,
the three-atom interaction potential is recovered.

The numerical calculations performed for medium-sized spheres show
that---compared to the case of the atoms being in free space---the
interatomic vdW interaction can be enhanced as well as reduced,
depending on the electromagnetical properties of the sphere, the
positions of the atoms relative to the sphere, and the positions of
the atoms relative to each other. 
In general, the electric properties
of the sphere have a more pronounced influence on the potential than
the magnetic properties.
We have
shown that 
the behavior of the interatomic potential can be 
qualitytively
understood
in the basis of an
image-charge model.

The results also indicate an essential difference between finite- and
infinite-sized systems, which particularly becomes apparent in the
linear configuration with a purely electrically polarizable sphere.
Depending on the distance between the surface and the neighbouring
atom, the (normalized) potential here approach different values,
while in the case of a half-space they converge to a single value.

\acknowledgments
This work was supported by the Deutsche Forschungsgemeinschaft.
We are grateful to the Ministry of Science, Research, and
Technology of Iran (H.~S.), the Alexander von Humboldt Foundation
(H.~T.~D. and S.~Y.~B.), and the National Program for Basic Research
of Vietnam (H.~T.~D.) for financial support.


\begin{appendix}

\section{Derivation of Eqs.~(\ref{G1})--(\ref{G4})}
\label{grn-sc}

To perform the summations over $m$ and $p$ in Eq.~(\ref{gs}),
we begin with the case
$p$ $\!=$ $\!-1$ and evaluate the
first term in the square brackets. Using Eqs.~(\ref{M}) and
(\ref{position}),
we may write
\begin{multline}
\label{a1}
\hspace{-1ex}\sum_{m=0}^nC_{nm}
 \mathbf{M}_{nm,-1}(\mathbf{r}_A,k)\mathbf{M}_{nm,-1}(\mathbf{r}_B,k)
 =\frac{Q^{(0)}_n\mathbf{e}_{\theta_A} \mathbf{e}_{\theta_B}}
 {\sin\theta_A\sin\theta_B}
\\
\times\sum_{m=0}^nC_{nm}m^2\cos(m\pi)
 P_n^m(\cos\theta_A)P_n^m(\cos\theta_B),
\end{multline}
where
\begin{equation}
C_{nm}\equiv\frac{(n-m)!}{(n+m)!}(2-\delta_{0m}).
\end{equation}
Differentiating
the addition theorem for spherical harmonics
\begin{equation}
\label{i}
\sum_{m=0}^nC_{nm}P_n^m(\cos\theta)P_n^m(\cos\theta')\cos(m\phi)
 =P_n(\psi),
\end{equation}
where
\begin{equation}
\label{psi}
\psi=\cos\theta\cos\theta'+\sin\theta\sin\theta'\cos\phi
,
\end{equation}
twice with respect to $\phi$,
we  obtain
\begin{multline}
\label{D2}
\sum_{m=0}^nC_{nm}
 m^2\cos(m\phi)P_n^m(\cos\theta)P_n^m(\cos\theta')
\\
=-\frac{{\mathrm d}^2 P_n(\psi)}{{\mathrm d}
 \psi^2}\biggl(\frac{{\mathrm d}\psi}{{\mathrm d}\phi}\biggr)^2
 -\frac{{\mathrm d}
 P_n(\psi)}{{\mathrm d}\psi}
 \frac{{\mathrm d}^2\psi}{{\mathrm d}
 \phi^2}.
\end{multline}
Using Eq.~(\ref{D2}) together with Eq.~(\ref{psi}) in
Eq.~(\ref{a1}), we find that
\begin{multline}
\label{sum-m-o}
\sum_{m=0}^nC_{nm}
 \mathbf{M}_{nm,-1}
 (\mathbf{r}_A,k)\mathbf{M}_{nm,-1}(\mathbf{r}_B,k)
\\
=-Q^{(0)}_n\,\frac{\mathrm{d}P_n(\gamma)}
 {\mathrm{d}\gamma}\,\mathbf{e}_{\theta_A} \mathbf{e}_{\theta_B}
\end{multline}
(recall that $\gamma$ $\!=$ $\!\cos\Theta$,
$\Theta$ $\!=$ $\!\theta_A$ $\!+$ $\!\theta_B$).
The summation over $m$
in the other terms in Eq.~(\ref{gs})
can be performed
in a similar way to obtain
\begin{multline}
\label{sum-m-e}
\sum_{m=0}^nC_{nm}
 \mathbf{M}_{nm,+1}(\mathbf{r}_A,k)\mathbf{M}_{nm,+1}(\mathbf{r}_B,k)
\\
=Q^{(0)}_nF_n(\gamma)\mathbf{e}_{\phi_A}
\mathbf{e}_{\phi_B},
\end{multline}
\begin{multline}
\label{sum-n-o}
\sum_{m=0}^nC_{nm}
 \mathbf{N}_{nm,-1}(\mathbf{r}_A,k)\mathbf{N}_{nm,-1}
 (\mathbf{r}_B,k)
\\
=-\frac{c^2}{\omega^2r_Ar_B}Q^{(3)}_n
 P'_n(\gamma)\mathbf{e}_{\phi_A} \mathbf{e}_{\phi_B},
\end{multline}
\begin{multline}
 \label{sum-n-e}
\sum_{m=0}^nC_{nm}
 \mathbf{N}_{nm,+1}(\mathbf{r}_A,k)\mathbf{N}_{nm,+1}
 (\mathbf{r}_B,k)
\\
=\frac{c^2}{\omega^2r_Ar_B}\Bigl\{[n(n+1)]^2Q_n^{(1)}
 P_n(\gamma)\mathbf
 e_{r_A}\mathbf
 e_{r_B}
\\
+n(n+1)\sin\Theta P'_n(\gamma)\left[Q_n^{(2)}
 \mathbf e_{r_A}\mathbf
 e_{\theta_B}+Q_n^{(3)}\mathbf
 e_{\theta_A}\mathbf
 e_{r_B}\right]
\\
-Q_n^{(4)}F_n(\gamma)\mathbf e_{\theta_A}
 \mathbf e_{\theta_B}\Bigr\}.
\end{multline}
Inserting
Eqs.~(\ref{sum-m-o})--(\ref{sum-n-e}) in Eq.~(\ref{gs}), we
arrive at Eqs.~(\ref{G1})--(\ref{G4}).


\section{The potential contributions
$\bm{U_{ij}^{(1)}}$, Eq.(\ref{vdw12}),
and $\bm{U_{ij}^{(1)}}$, Eq.(\ref{vdw22})}
\label{Uij}

We substitute Eqs.~(\ref{G6})--(\ref{G8}) together with
Eqs.~(\ref{G1})--(\ref{G4}) into Eqs.~(\ref{vdw12}) and (\ref{vdw22})
and obtain the following expressions for the nonzero
$U^{(1)}_{ij}$ and $U^{(2)}_{ij}$:
\begin{multline}
\label{u1rr}
U^{(1)}_{rr}(\mathbf{r}_A,\mathbf{r}_B)=
 -\frac{\hbar\mu_0^2c^3}{16 \pi^3l^5r_Ar_B}
 \sum_{n=1}^\infty n(n+1)(2n+1)
  P_n(\gamma)
\\
\times\int_0^\infty{\mathrm{d}}u\,u\,e^{-lu/c}\alpha_A(iu)\alpha_B(iu)
 B_n^N(iu)Q_n^{(1)}
\\
\times \left[l^2f(\xi)\cos\Theta
-g(\xi)l_Al_B\right],
\end{multline}
\begin{multline}
\label{u1rt}
U^{(1)}_{
\genfrac{}{}{0pt}{}{
r\theta
}
{
\theta r
}}
(\mathbf{r}_A,\mathbf{r}_B)=
 - \frac{\hbar\mu_0^2c^3\sin^2\Theta}{16\pi^3l^5r_Ar_B}
 \sum_{n=1}^\infty(2n+1)P_n'(\gamma)
\\
 \times\int_0^\infty{\mathrm{d}}u\,u\,e^{-lu/c}
 \alpha_A(iu)\alpha_B(iu)
  B_n^N(iu)
  \genfrac{}{}{0pt}{}{Q_n^{(2)}}{Q_n^{(3)}}
\\
\times \left[l^2f(\xi)\pm g(\xi)r_{A(B)} l_{A(B)}\right] ,
\end{multline}
\begin{multline}
\label{u1tt}
U^{(1)}_{\theta\theta}(\mathbf{r}_A,\mathbf{r}_B)=
 \frac{\hbar\mu_0^2c}{16\pi^3l^5}
 \sum_{n=1}^\infty
 \frac{(2n+1)}{n(n+1)}\int_0^\infty{\mathrm{d}}u\,u^3
 \,e^{-lu/c}
\\
\times\!\alpha_A(iu)\alpha_B(iu)
\biggl[B_n^M(iu)Q_n^{(1)}P'_n(\gamma)
-\frac{c^2B_n^N(iu)}{u^2r_Ar_B}
\\
\times Q_n^{(4)}F_n(\gamma)
 \biggr]\!\left[l^2f(\xi)\cos\Theta-g(\xi)r_Ar_B\sin^2\Theta\right],
\end{multline}
\begin{multline}
\label{u1ff}
U^{(1)}_{\phi\phi}(\mathbf{r}_A,\mathbf{r}_B)=
 \frac{\hbar\mu_0^2c}{16\pi^3l^3}
 \sum_{n=1}^\infty
 \frac{(2n+1)}{n(n+1)}\int_0^\infty{\mathrm{d}}u\,u^3
 \,e^{-lu/c}
\\
\times\!\alpha_A(iu)\alpha_B(iu)\biggl[B_n^M(iu)Q_n^{(1)} F_n(\gamma)
\\
-\frac{c^2B_n^N(iu)}{u^2r_Ar_B}\,
Q_n^{(4)}P_n'(\gamma)
 \biggr] f(\xi),
\end{multline}
\begin{multline}
\label{u2rr}
U^{(2)}_{rr}(\mathbf{r}_A,\mathbf{r}_B)\!=\!
 \frac{-\hbar\mu_0^2c^2}{32\pi^3r_A^2r_B^2}\!
 \sum_{m,n=1}^\infty
 m(m\!+\!1)(2m\!+\!1)
\\
\times n(n\!+\!1)(2n\!+\!1)P_m(\gamma)P_n(\gamma)
 \int_0^\infty{\mathrm{d}}u\,u^2
\\
\times
 \alpha_A(iu)\alpha_B(iu)B_m^N(iu)B_n^N(iu)Q_m^{(1)}Q_n^{(1)},
\end{multline}
\begin{multline}
\label{u2rt}
U^{(2)}_{
\genfrac{}{}{0pt}{}{
r\theta
}
{
\theta r
}}
(\mathbf{r}_A,\mathbf{r}_B)\!=\!
\frac{-\hbar\mu_0^2c^2\sin^2\Theta}{32\pi^3r_A^2r_B^2}\!
\sum_{m,n=1}^\infty
(2m\!+\!1)(2n\!+\!1)
\\
\times\! P_m'(\gamma)P_n'(\gamma)
 \int_0^\infty{\mathrm{d}}u\,u^2\alpha_A(iu)\alpha_B(iu)
\\
\times  B_m^N(iu)B_n^N(iu)
\genfrac{}{}{0pt}{}{Q_m^{(2)}Q_n^{(2)}}{Q_m^{(3)}Q_n^{(3)}},
\end{multline}
\begin{multline}
\label{u2tt}
U^{(2)}_{
\genfrac{}{}{0pt}{}
{
\theta \theta
}
{
\phi\phi
}}
(\mathbf{r}_A,\mathbf{r}_B)=
\frac{-\hbar\mu_0^2}{32\pi^3c^2}\sum_{m,n=1}^\infty
 \frac{(2n\!+\!1)}{n(n\!+\!1)}\frac{(2m\!+\!1)}{m(m\!+\!1)}
\\
\times\int_0^\infty{\mathrm{d}}u\,u^6\alpha_A(iu)\alpha_B(iu)
 \biggl[B_m^MQ_m^{(1)}
\genfrac{}{}{0pt}{}{P'_m(\gamma)}{F_m(\gamma)}
\\
-\frac{c^2B_m^N}{u^2r_Ar_B} Q_m^{(4)}
\genfrac{}{}{0pt}{}{F_m(\gamma)}{P_m'(\gamma)}
\biggr]
\biggl[B_n^MQ_n^{(1)}
\genfrac{}{}{0pt}{}{P'_n(\gamma)}{F_n(\gamma)}
\\
-\frac{c^2B_n^N}{u^2r_Ar_B} Q_n^{(4)}
\genfrac{}{}{0pt}{}{F_n(\gamma)}{P_n'(\gamma)}
\biggr].
\end{multline}


\section{The limiting cases of a large and a small sphere}
\label{ls}

When in the case of a large sphere the conditions~(\ref{dr})
and (\ref{theta}) are satisfied, then the leading
contributions to the sums in
Eqs.~(\ref{G1})--(\ref{G4}) come from terms with
\mbox{$n$ $\!\gg$ $\!1$}
(also see Ref.~\cite{st1}),
for which the spherical Bessel and Hankel functions
approximate to \cite{abr}
\begin{equation}
\label{j1}
j_n(z)
=
\frac{z^n}{(2n+1)!!}
\left[1-\frac{z^2}{4n+6}+\frac{z^4}{(16n+24)(2n+5)}\right]
\end{equation}
and
\begin{equation}
\label{h1}
h_n^{(1)}(z)
=
-i\frac{(2n-1)!!}{z^{n+1}}
\!\left[1\!+\!\frac{z^2}{4n-2}\!+\!\frac{z^4}{(16n-8)(2n-3)}
\right]\!,
\end{equation}
respectively. Hence,
Eqs.~(\ref{BM}) and (\ref{BN})
approximate to
\begin{equation}
\label{BM3}
B_n^M(iu)
= \frac{i(iRu/c)^{2n+1}}{16n[(2n+1)!!]^2[\mu(iu)+1]}
\left(a_2n^2\!+\!a_1n\!+\! a_0\right)
\end{equation}
and
\begin{equation}
\label{BN3}
B_n^N(iu)
= \frac{i(iRu/c)^{2n+1}}{16n[(2n+1)!!]^2[\varepsilon(iu)+1]}
\left(b_2n^2\!+\!b_1n\!+\! b_0\right)
\!,
\end{equation}
respectively, where
\begin{multline}
\label{a_0}
a_0=\left\{488+[1+\varepsilon(iu)\mu(iu)]^2
 R^4u^4/c^4\right\}[\mu(iu)-1]
\\
-8\left\{7\mu(iu)[\varepsilon(iu)\!-\!1]\!-\!
 5\varepsilon(iu)\mu(iu)^2\!+\!5\right\}R^2u^2/c^2,
\end{multline}
\begin{align}
\label{a_1}
a_1&= 8
 \left\{26+[1+\varepsilon(iu)\mu(iu)]R^2u^2/c^2\right\}[\mu(iu)-1]
\nonumber\\
&\equiv 8 \eta(iu)[\mu(iu)-1],
\end{align}
\begin{equation}
\label{a_2}
a_2=32[\mu(iu)-1],
\end{equation}%
and $b_0$, $b_1$, and $b_2$ can be found from $a_0$, $a_1$, and $a_2$,
respectively, by interchanging $\mu$ and $\varepsilon$.
Equations~(\ref{Q1})--(\ref{Q4}) then approximate to
\begin{equation}
\label{q1}
Q_n^{(1)}=-\left(\frac{ic}{u}\right)^{2n+2}
\frac{[(2n-1)!!]^2}{(r_Ar_B)^{n+1}},
\end{equation}
\begin{equation}
\label{q123}
Q_n^{(2)}=Q_n^{(3)}=-Q_n^{(4)}/n=-nQ_n^{(1)}.
\end{equation}

In order to illustrate
the application of the approximation scheme
to the Green tensor elements~(\ref{G1})--(\ref{G4}), let us
consider the element
$G_{rr}^{(1)}(\mathbf{r}_A,\mathbf{r}_B ,iu)$. Inserting
Eqs.~(\ref{BN3}) and (\ref{q1}) in
Eq.~(\ref{G1}), we find
\begin{equation}
\label{Grr-l}
G_{rr}^{(1)}(\mathbf{r}_A,\mathbf{r_B},iu)
 =\frac{-c^2t}{128\pi
  u^2R^3[\varepsilon(iu)+1]}\sum_{k=0}^2a_ks_k,
\end{equation}
where
\begin{equation}
s_k=\sum_{n=1}^\infty n^k t^{n+1}P_n(\gamma)
\end{equation}%
and $t$ $\!=$ $\!R^2/(r_Ar_B)$. Differentiating
the identity
\begin{equation}
\sum_{n=1}^\infty t^n
 P_n(\gamma)=\frac{1}{\sqrt{1-2t\gamma+t^2}}-1
\end{equation}
\cite{abr} with respect to $t$, we can perform
the summation in Eq.~(\ref{Grr-l}) to obtain
\begin{equation}
\label{e74}
s_0=\frac{t}{\sqrt{1-2t \gamma+t^2}}-t,
\end{equation}
\begin{equation}
\label{e74-1}
s_1= \frac{t^2\gamma-t^3}{\left[1-2t \gamma+t^2\right]^{3/2}}
\end{equation}
\begin{equation}
\label{e74-2}
s_2=\frac{3\gamma^2-4\gamma t+2t^2-1}
{\left[1-2t \gamma+t^2\right]^{5/2}}
\end{equation}%

Recalling the conditions~(\ref{dr}) and (\ref{theta}),
we can further simplify the result. Up to second order in the
small parameters $\delta_{A'}/R$, we have
\begin{equation}
\label{e75}
t^k
=
1-k\frac{\delta_A+\delta_B}{R}+\frac{k(k+1)}{2}
\frac{\delta_A^2+\delta_B^2}{R^2}+k^2\frac{\delta_A\delta_B}{R^2},
\end{equation}
implying that
\begin{equation}
\label{e76}
1-2t \gamma+t^2
\simeq
\Theta^2 +\frac{(\delta_A+\delta_B)^2}{R^2}=\frac{l_+^2}{R^2}.
\end{equation}
Using Eqs. (\ref{e74})--(\ref{e76}) in Eq.~(\ref{Grr-l}), we
find that within this order,
\begin{multline}
\label{Grr-h}
G_{rr}^{(1)}(\mathbf{r}_A,\mathbf{r_B},iu)
 =\frac{c^2}{128\pi
 u^2l_+^5[\varepsilon(iu)+1]}
\\
 \times\bigg\{32(X^2-2\delta_+^2)[\varepsilon(iu)-1]
 -8\frac{\delta_+l_+^2}{R}\eta(iu)[\varepsilon(iu)-1]
\\
 -\bigg
[
\bigg
(
488+[1+\varepsilon(iu)\mu(iu)]^2
\frac{R^4u^4}{c^4}\bigg
)
[\varepsilon(iu)-1]
\\
-8\Big[7\varepsilon(iu)[\mu(iu)\!-\!1]\!-\!
5\varepsilon^2(iu)\mu(iu)\!+\!5\Big]\frac{R^2u^2}{c^2}\bigg
]
\frac{l_+^4}{R^2}\bigg\}.
\end{multline}
Recalling
that
$X$, $l_+$, $\delta_+$ $\!\ll$ $\!R$, it can be seen that unless
\mbox{$|\varepsilon(iu)$ $\!-$ $\!1|$ $\!\ll$ $\! 1$}, the 
third term in the curly bracket
in Eq.~(\ref{Grr-h}) can be
approximately
ignored.
Hence, 
\begin{multline}
\label{Grr-h1}
G_{rr}^{(1)}(\mathbf{r}_A,\mathbf{r_B},iu)
 =\frac{c^2}{4\pi
 u^2l_+^5}\frac{\varepsilon(iu)-1}{\varepsilon(iu)+1}
\\
 \times\left\{
 (X^2-2\delta_+^2)
 -\frac{\delta_+l_+^2}{16R}
 \left[26+\varepsilon(iu)\mu(iu)R^2u^2/c^2\right]
 \right\}\!.
\end{multline}
In the case of a purely magnetically polarizable sphere
[$\varepsilon(iu)$ $\!=$ $\!1$], the leading contribution
to $G_{rr}^{(1)}$ comes from the third term in the
curly brackets in Eq.~(\ref{Grr-h}):  
\begin{equation}
\label{Grr-h2}
G_{rr}^{(1)}(\mathbf{r}_A,\mathbf{r_B},iu)
 =\frac{\mu(iu)-1}{16\pi l_+}.
\end{equation}

The other
Green tensor
elements
can be evaluated
in a quite similar way. Substituting the resulting
expressions into Eqs.~(\ref{vdw12}) and (\ref{vdw22}),
and summing them in accordance with
Eq.~(\ref{us2}), we eventually arrive
at Eqs.~(\ref{l-e}) and (\ref{l-m}).

In the limiting case of a small sphere where
the condition~(\ref{smalls}) holds, the leading
contributions to the frequency integrals in
Eqs.~(\ref{u1rr})--(\ref{u2tt})
come from the region where
$u$ $\!\ll$ $\!c/R$,
or equivalently \mbox{$|z_0|$, $|z_1|$
$\!\ll$ $\! 1$}
(also see Ref.~\cite{st1}).
In this region we may approximate the spherical
Bessel and Hankel functions appearing in Eqs.~(\ref{BM}) and
(\ref{BN}) by their next-to-leading order expansions in $z$
\cite{abr}, i.e.,
\begin{equation}
\label{j}
j_n(z)
=
\frac{z^n}{(2n+1)!!}\left[1-\frac{z^2}{4n+6}\right]
\end{equation}
and
\begin{equation}
\label{h}
h_n^{(1)}(z)
=
-i\frac{(2n-1)!!}{z^{n+1}}
 \left[1-\frac{z^2}{2-4n}\right],
\end{equation}
so that
Eqs.~(\ref{BM}) and (\ref{BN}),
respectively,
approximate to
\begin{multline}
\label{BM1}
B_n^M(iu)= i\frac{2n+1}{[(2n+1)!!]^2[\mu(iu)n+n+1]}
\\
\times\left\{[\mu(iu)-1](n+1)
\left(\frac{iRu}{c}\right)^{\!2n+1}
 +O\!\left(\frac{iRu}{c}\right)^{\!2n+3}\right\}
\end{multline}
and
\begin{multline}
\label{BN1}
B_n^N(iu)= i\frac{2n+1}{[(2n+1)!!]^2[\varepsilon(iu)n+n+1]}
\\
\times\left\{[\varepsilon(iu)-1](n+1)
 \left(\frac{iRu}{c}\right)^{\!2n+1}
 +O\!\left(\frac{iRu}{c}\right)^{\!2n+3}\right\},
\end{multline}
revealing that
in Eq.~(\ref{us2})
the $U_{ij}^{(2)}$ terms are small
in comparison to
the $U_{ij}^{(1)}$ terms and can be neglected,
so that,
in
leading order of $Ru/c$,
\begin{equation}
\label{uija}
U^{(b)}
=
\sum_{i,j=r,\theta,\phi}U^{(1)}_{ij}
 (\mathbf{r}_A,\mathbf{r}_B).
\end{equation}
Further, it can be seen that
in the sums in Eqs.~(\ref{G1})--(\ref{G4})
the terms
with $n$ $\!=$ $\!1$
are the leading ones, for which
\begin{equation}
\label{q11}
Q_1^{(1)}=-\frac{(1+a)(1+b)}{a^2b^2}\,e^{-a-b},
\end{equation}
\begin{equation}
\label{q2}
Q_1^{(2)}=\frac{(1+a)f(b)}{a^2b^2}\,e^{-a-b},
\end{equation}

\begin{equation}
\label{q3}
Q_1^{(3)}=\frac{f(a)(1+b)}{a^2b^2}\,e^{-a-b},
\end{equation}
\begin{equation}
\label{q4}
Q_1^{(4)}=-\frac{f(a)f(b)}{a^2b^2}\,e^{-a-b},
\end{equation}
\begin{equation}
\label{ffff}
F_1(\gamma)=P_1(\gamma)=\gamma,
\end{equation}
\begin{equation}
\label{BM4}
B_1^M(iu)=\frac{2}{3}\frac{\mu(iu)-
 1}{\mu(iu)+2}\left(\frac{Ru}{c}\right)^3,
\end{equation}
\begin{equation}
\label{BN4}
B_1^N(iu)=\frac{2}{3}\frac{\varepsilon(iu)-
 1}{\varepsilon(iu)+2}\left(\frac{Ru}{c}\right)^3.
\end{equation}
Substituting Eqs.~(\ref{u1rr})--(\ref{u1ff}) together with
Eqs.~(\ref{q11})--(\ref{BN4}) in Eq.~(\ref{uija}), 
we arrive at Eq.~(\ref{s}).
\end{appendix}


\end{document}